\documentclass[prl,twocolumn,amsmath,amssymb,showpacs]{revtex4}

\usepackage{graphicx}% Include figure files
\usepackage{dcolumn}% Align table columns on decimal point
\usepackage{bm}% bold math%\nofiles%A. Perrin, H. Chang, V. Krachmalnicoff, D. Boiron, A. Aspect & C. I. Westbrook

\begin{document}%\preprint{APS/123-QED}
\title{Observation of atom pairs in spontaneous four wave mixing of two colliding Bose-Einstein Condensates}
\author{A. Perrin, H. Chang, V. Krachmalnicoff, M. Schellekens, D. Boiron, A. Aspect and C. I. Westbrook} 
\email{christoph.westbrook@institutoptique.fr}%
%\homepage{http://www.atomoptic.fr}
\affiliation{Laboratoire Charles Fabry de l'Institut d'Optique, CNRS, Univ Paris-Sud\\Campus Polytechnique, RD128, 91127 Palaiseau cedex }%\author{Charlie Author} %%\homepage{http://www.Second.institution.edu/~Charlie.Author}%\affiliation{%Second institution and/or address\\%This line break forced% with \\%}%
\date{\today}% It is always \today, today,             %  but any date may be explicitly specified
\begin{abstract}
We study atom scattering from two colliding Bose-Einstein condensates using a position sensitive, time resolved, single atom detector. In analogy to quantum optics, the process can also be thought of as spontaneous, degenerate four wave mixing of de Broglie waves. We find a clear correlation between atoms with opposite momenta, demonstrating pair production in the scattering process. We also observe a  Hanbury Brown and Twiss correlation for collinear momenta, which permits an independent measurement of the size of the pair production source and thus the size of the spatial mode. 
The back to back pairs occupy very nearly two oppositely directed spatial modes, a promising feature for future quantum optics experiments.
\end{abstract}% PACS, the Physics and Astronomy Classification Scheme.
%\keywords{Suggested keywords}%Use showkeys class option if keyword display desired
\maketitle

Recent years have seen the emergence of  ``quantum atom optics", that is the extension of the many analogies between atom optics and traditional optics to the quantum optical domain in which phenomena like vacuum fluctuations and entanglement play a central role. In optics the advent of correlated photon pairs \cite{burnham:70} has provided a fruitful avenue of investigation, with examples including single photon sources and entangled states \cite{walls:91}. Partly inspired by this work, there have been many proposals concerning atom pairs, especially the production and observation of entanglement \cite{pu:00,duan:00,kheruntsyan:05,opatrny:01,savage:06}. Many authors have also theoretically investigated other aspects of the pair production mechanism in both atomic collisions and in the breakup of diatomic molecules \cite{band:00,naidon:03,zin:05,zin:06,savage:06,norrie:06,deuar:07}.

As emphasized in Ref.~\cite{duan:00}, pair production can be studied in two limits. If many atoms are created in a single mode, stimulated emission of atoms is important, and one can speak of two mode squeezing in analogy with Ref.~\cite{heidmann:87}. The opposite limit, in which the occupation number of the modes is much less than unity, corresponds to the spontaneous production of atom pairs, entangled either in spin or momentum in analogy with Ref.~\cite{ou:88,rarity:90}. Experiments on stimulated atomic four wave mixing \cite{deng:99, vogels:02, vogels:03} and on parametric amplification in an optical lattice \cite{gemelke:05,campbell:06} are in the first limit, and  pairs of  ``daughter BEC's" with opposite velocities have been clearly observed. Experiments in the regime of individual atom pairs include the many experiments investigating the scattered halo in collisions of cold atoms either in the s-wave regime \cite{chikkatur:00,gibble:95,katz:05} or for higher partial waves \cite{buggle:04,thomas:04}. None of these experiments however, has demonstrated correlated pairs.  The only evidence of atom pair production with cold atoms has been reported in absorption images of atoms from the breakup of molecules near a Feshbach resonance \cite{greiner:05}.
%and even this experiment did not undertake a quantitative study of the process.

\begin{figure}[ht!]
\includegraphics[width=\linewidth]{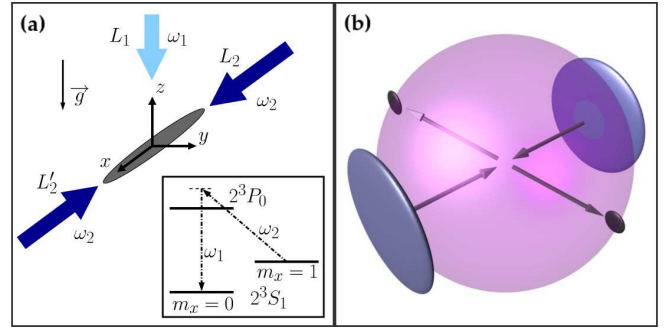}
\caption{\label{fig:exp}{(Color online) (a) View of the magnetically trapped condensate (in the $m_x=1$ state) and the three laser beams which create two cigar shaped counterpropagating free condensates (in the $m_x=0$ state) by $\sigma^-/\pi$ Raman transfers induced by $L_1-L_2$ and $L_1-L'_2$ respectively (see inset). $L_1$ is $\pi-$polarized (along $x$) while $L_2$ and $L'_2$ are $\sigma^-$-polarized.  (b) Representation in velocity space of the expected atomic density after the collision. The scattered atoms are on a sphere and the remaining condensates,  pancake-shaped after expansion, lie on the edge of the sphere along the $x$ axis.}}
\end{figure}

Here, we report on the observation of individual atom pairs with opposite velocities produced in the collision of two condensates.
A time and position resolved, single atom detector \cite{schellekens:05} permits us to reconstruct the 3 dimensional distribution of the scattered atoms:  a spherical shell in velocity space. We also reconstruct the two-particle correlation function in 3D and find a strong correlation between atoms emitted back to back. This process can be interpreted as a  spontaneous four wave mixing process constrained by a phase matching condition as in the non linear optical analog which produces twin photons \cite{walls:91}.
%\cite{scully:97}. 
It can also be seen as the result of pairwise elastic collisions between atoms, constrained by momentum conservation.  We measure the width of the velocity correlation function for a back to back atom pair and show that it can be roughly understood from the uncertainty limited momentum spread of the colliding BECs. 

This interpretation is confirmed by the observation of the velocity correlation function for two atoms scattered in the {\it same} direction. This latter effect, predicted in Refs.~\cite{zin:05,savage:06,deuar:07}, is another manifestation of the Hanbury Brown-Twiss effect (HBT). As in high energy collisions \cite{baym:98}, the effect allows us to measure the size of the collision volume. 

The fact that the width of the HBT peak is close to that of the back to back correlation confirms that for a given atom on the collision sphere, its partner is scattered into a single mode of the matter wave field. This observation is crucial for future experiments in which one would like to bring pairs back together in order to confirm their entanglement in the spirit of Ref.~\cite{rarity:90} or observe other quantum effects \cite{hong:87}.

We produce condensates of $10^4 - 10^5$ atoms in the $m_x=1$ sublevel of the $2^3S_1$ state of metastable helium (He*). The condensates are stored in a cylindrically symmetric magnetic trap with axial and radial trapping frequencies of $47$~Hz and $1150$~Hz respectively. The bias field is 0.25 G in the $x$ direction (see Fig. \ref{fig:exp}), and defines the quantization axis. The uncertainty limited velocity spread of the colliding atoms is thus anisotropic and we calculate it numerically using the Gross-Pitaevskii equation \cite{epaps:07}. For a condensate with  $3\times 10^4$ atoms, we find rms axial and radial velocity spreads of $v_{x}^{\rm rms}=0.0044\,v_{\rm rec}$ and $v_{yz}^{\rm rms}=0.091\,v_{\rm rec}$,
where $v_{\rm rec}=9.2$~cm/s is the single photon recoil velocity, 
$\hbar k /m$ where $k$ is the photon wavevector and $m$ is the atomic mass.
The spread in these values due to the spread in condensate number is about $\pm 20~\%$.

To generate two colliding Bose-Einstein condensates, we use two stimulated Raman transitions with different momentum transfers, produced by phase coherent laser beams $L_1$, $L_2$ and $L^\prime_2$, as shown in Fig.~\ref{fig:exp} \cite{epaps:07}. These transitions have two purposes: first they transfer atoms to the magnetic field insensitive state $m_x=0$ so that they freely fall to the detector and second, they separate the condensate into two components with velocities $v_{\rm rec}(\mathbf{e_1} \pm \mathbf{e_2})$, where $\mathbf{e_1}$ and  $\mathbf{e_2}$ are the unit vectors along the propagation axes of the laser beams $L_1$ and $L_2$ respectively. The beams are pulsed on for a duration of $\sim 500$~ns and couple about $60~\%$ of the atoms to the $m_x=0$ state. We do not switch off the magnetic trap, therefore atoms remaining in $m_x=1$ stay trapped. The two colliding condensates travel with a relative velocity of $2 v_{\rm rec}$, at least 8 times larger than the speed of sound 
 in the initial condensate. This ensures that elementary excitations of the condensate correspond to free particles. Since they are no longer trapped, the two colliding condensates expand radially, reducing the collision rate. A numerical model \cite{band:00}, assuming an expansion identical to that of a single condensate with the same total number of atoms, shows a roughly exponential decrease in the pair production rate with a time constant of  $\sim150\, \mu$s.

\begin{figure}[htb]
\includegraphics[width=\linewidth]{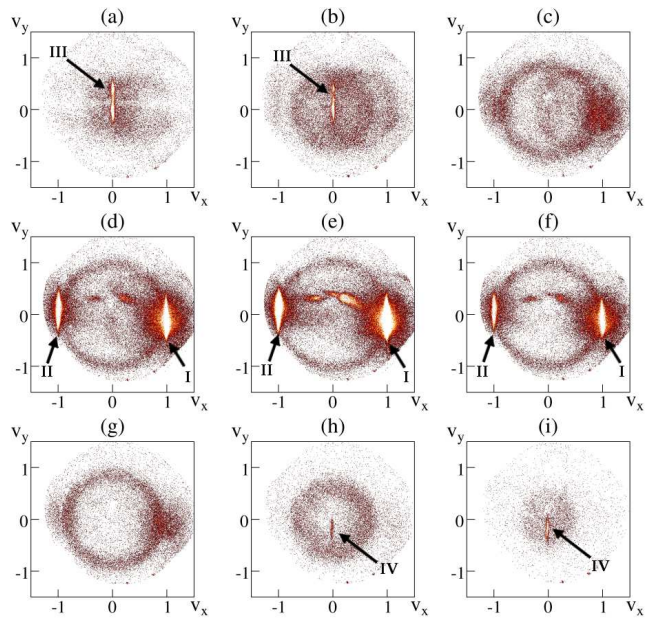}
\caption{\label{fig:sphere}(Color online) (a-i) Images of the collision of two condensates. Each frame represents a $2.4$ ms time slice of the atomic cloud as it passes the plane of the detector ($x-y$). 150 shots have been averaged to obtain these images. The two colliding condensates I, II and the collision sphere are clearly visible. Other features visible in the images are discussed in the text. The axes are marked in units of the recoil velocity.}
% The size of each image is $8\,{\rm cm} \times 8\,{\rm cm}$
\end{figure}

After the collision, atoms fall onto a 8 cm microchannel plate detector placed 46.5 cm below the trap center. This detector measures the arrival time of the atoms and their positions in the $x-y$ plane \cite{schellekens:05, jeltes:07}. Figure \ref{fig:sphere} shows successive 2.4~ms time slices showing the atom positions as they cross the detector plane. The time of flight for the center of mass to reach the detector is $320$~ms. Since this time of flight is large compared to the collision duration, and the observed patterns are large compared to the collision volume, the observed 3D atom positions accurately reflect the velocity distribution after collision. In the following, we will only refer to the velocities of the detected atoms.

In Fig.~\ref{fig:sphere}, one clearly sees a spherical shell of radius of $v_\text{rec}$, represented by circles of varying diameter. In the mid plane of the sphere one can see the unscattered, pancake-shaped condensates I, II which locally saturate the detector. Other features are also visible in Fig.~\ref{fig:sphere}. In frames (a, b) one sees a condensate, III, which underwent no momentum transfer, possibly due to the imperfect polarization of the Raman beams which can produce an off resonant, single beam Raman transition \cite{epaps:07}. A fourth condensate, IV,  probably resulting from four-wave mixing \cite{deng:99} of condensate III and the main unscattered condensates I, II is visible in frames (h,i).  Frames (b,c) show a collision sphere due to the collision of I with atoms remaining trapped in $m_x=1$ and with condensate III. The two spots within the sphere in frames (d-f) are not understood.

To avoid effects of local saturation of the detector in our analysis, we exclude regions around the 4 condensates, representing about $40~\%$ of the sphere. On the remaining area of the sphere we detect between 30 and 300 atoms on each shot, with an average of about 100 per shot. Assuming a detection efficiency of 10~\% \cite{jeltes:07}, this means that $\sim 5$~\% of the atoms are scattered from the two condensates. This number is consistent with the expected $s-$wave cross section \cite{scatlength:07} and the estimated evolution of the density during the collision.

\begin{figure}[t]
\includegraphics[width=\linewidth]{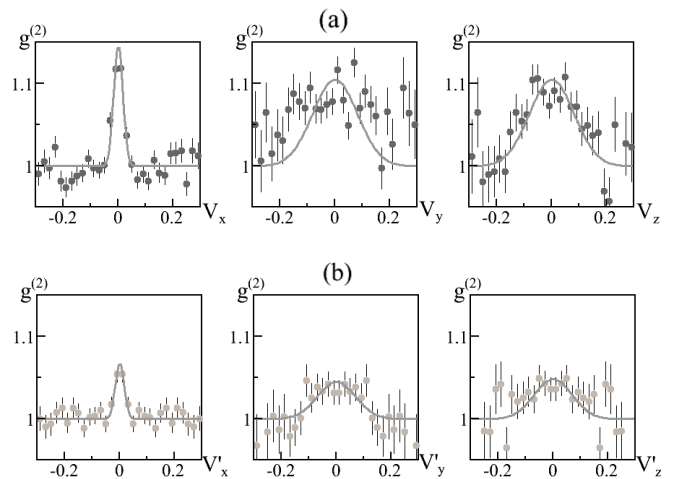}
\caption{\label{fig:cor} Back to back (panel a) and collinear (panel b) correlation peaks. (a) Projection of $g^{(2)}(\mathbf{V}=\mathbf{V_1}+\mathbf{V_2})$ along the different axes of the experiment and around $\mathbf{V}=\mathbf{0}$ . The projection consists in averaging the correlation in the two other directions over a surface equal to the products of the corresponding correlation lengths. This averaging makes the height smaller than the 3D fitted value $\eta^{\rm BB}=0.19\pm0.02$. The peak is the signature for correlated atoms with opposite velocities. 
%The projection of the 3D fit (see text) is also showed in the graphs. Along the $y$ and $z$ axis the thickness of the scattering shell, $\sim 0.08 v_{\rm rec}$, is a limiting factor. The signal to noise ratio decreases fast outside this zone. This can explain why some points in these graphs are off the error bars. 
(b) Projection of $g^{(2)}(\mathbf{V'}=\mathbf{V_1}-\mathbf{V_2})$ along the different axes of the experiment. This peak is due to the Hanbury Brown and Twiss bunching effect. All velocities are expressed in units of the recoil velocity.}
\end{figure}

We examine the pair correlation function for atoms in back to back directions by constructing, within the set of all the scattered atoms in one shot, a three dimensional histogram containing all the pairs with a velocity sum ${\bf V}=\mathbf{V_1}+\mathbf{V_2}$ close to zero. We then sum the histograms over $1100$ shots. Another histogram containing all the pairs of the sum of all shots gives the accidental coincidence rate for uncorrelated atoms and is used as a normalization. We thus recover the normalized second order correlation function, averaged over the sphere \cite{epaps:07}, $g^{(2)}(\bf V)$ of the distribution of relative velocities of atom pairs on the sphere. Figure \ref{fig:cor}(a) shows the behavior of $g^{(2)}(\bf V)$ around $\bf V=0$ projected along the three space axes. The peak indicates that, given the detection of an atom on the sphere, there is an enhanced probability of detecting a second one on the opposite side. Cartesian coordinates are best suited to plotting the data because of the competing spherical symmetry of the scattering process and the cylindrical symmetry of the source.

To analyze these results further, we perform a three-dimensional Gaussian fit to the normalized histogram : 
\begin{equation}
g^{(2)}(V_x,V_y,V_z)=1+\eta \ e^{-\frac{{V_x}^2}{2{\sigma_x}^2}-\frac{V_y^2+V_z^2}{2{\sigma_{yz}}^2}}. 
\label{fit}
\end{equation}
The fit gives $\eta^{\rm BB}=0.19\pm0.02$, $\sigma^{\rm BB}_x=0.017\pm 0.002\, v_{\rm rec}$ and $\sigma^{\rm BB}_{yz}=0.081\pm0.004 \, v_{\rm rec}$. 
The observed width in the $x$ direction is limited by the rms pair resolution of the detector, 
$0.014\, v_{\rm rec}$ \cite{gomes:06,epaps:07}.
In the $y$ and $z$ directions, the observed width is close to the uncertainty limited 
velocity scale $v_{yz}^{\rm rms}$ discussed above.  
It is therefore reasonable to conclude that the anisotropy in the correlation function is closely related
to the anisotropy of the momentum distribution in the source. Detailed modelling accounting accurately for this width is in progress, but for purposes of this letter,
we will simply compare the width with that of the correlation function for collinear atoms as described below. 
 
%$\spadesuit$ A simple way to explain the anisotropic broadening of the correlation peak comes from the initial momentum spread of the colliding condensates, which is reflected in the momentum spread of the center of mass momentum, which is the quantity measured here according to the classical model of elastic collisions between individual atoms with isotropic cross section.\emph{are sure you are plotting $V_1 +V_2$ and not half of it ?} .$\clubsuit$ In the $x$ direction the width is limited by  the single particle rms detector resolution $\clubsuit$, which has a rms width of $\delta = 0.012 \, v_{\rm rec}$.\emph{OK?}$\spadesuit$

 The procedure to construct the correlation function for nearly collinear velocities (the HBT effect) 
is the same as that for
the back to back correlation function. 
Defining the relative velocity ${\bf V'}=\mathbf{V_1}-\mathbf{V_2}$ 
we show in Fig.~\ref{fig:cor}{b} the correlation function $g^{(2)}(\bf V')$ around $\bf V'=0$. 
Using the fitting function, Eq.~\ref{fit}, we find: $\eta^{\rm CL}=0.10\pm0.02$, $\sigma^{\rm CL}_x=0.016\pm0.003 \, v_{\rm rec}$ and $\sigma^{\rm CL}_{yz}=0.069\pm0.008 \, v_{\rm rec}$.
As in the back to back case, the width in the $x$~direction is limited by the resolution while
in the $y-z$~plane it is close to $v_{yz}^{\rm rms}$. 
If we think of the HBT effect as giving a measure of the size of the pair production source, 
the width of the collinear correlation function defines the size of a 
mode of the scattered matter wave field.
The fact that the back to back and collinear widths are so close, at least in the
directions we can resolve, is further, strong evidence that, at least in the directions we resolve, the
pairs we produce are in oppositely directed modes.

We now turn to the height of the peaks $\eta$. In the collinear case we expect the value of $\eta^{\rm CL}$ to be unity for a detector resolution much smaller than the peak width. Since in the $x$ direction the width is clearly limited by the resolution, a crude estimate for $\eta^{\rm CL}$ is the ratio of the ideal width to the observed one: $\eta^{\rm CL}\approx v_{x}^{\rm rms}/\sigma_{x}=0.3$. The discrepancy with the fitted value may have to do with our crude estimate of the effective source size along $x$ and therefore of $v_{x}^{\rm rms}$.  

In the back to back case, the height of the peak is not limited to unity. A simple model of the peak height compares the number of true pairs to random coincidences in a volume $\Delta V$ defined by the widths observed in Fig.~\ref{fig:cor}:
\begin{equation}\label{eq:g2}
1+\eta^{\rm BB}=\frac{\rm true + random}{\rm random}=1+\frac{V}{N \Delta V}\end{equation}
Here $N$ is the number of atoms scattered on a single shot (but not necessarily detected) and $V$ is the volume of the scattering shell. A rough estimate of  $\Delta V/V$ is $1/1400$. As mentioned above, we detect on average 100 atoms on the analyzed 60~\% of the sphere.  Assuming again a quantum efficiency of 10~\%, a rough estimate of the average number $N$ is 1700 so that we find $\eta^{\rm BB} \approx 0.8$ which gives the correct order of magnitude. We emphasize that $\Delta V$ is limited by the detector resolution in the $x$ direction and is therefore about 10 times larger than the volume corresponding to a single mode. Thus as stated in the introduction, the number of scattered atoms per mode is small compared to unity, and we are in the separated entangled pair production regime. We can verify the $1/N$ dependence of Eq.~(\ref{eq:g2}) by binning the data according to the number of scattered atoms per shot. Dividing the 1100 shots into 3 bins of different atom numbers we do observe the expected trend as shown in Fig. 4. \begin{figure}[htb]
\includegraphics[width=\linewidth]{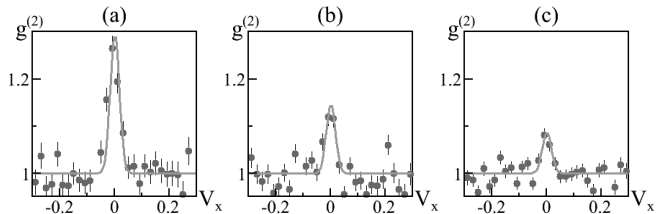}
\caption{\label{fig:corh} Projections of $g^{(2)}(\mathbf{V})$ along the $x$ axis and around $\mathbf{V}=\mathbf{0}$. Bin of mean number of detected atoms of (a) 50, (b) 125 and (c) 190. 
%The data thus confirm Eq.~\ref{eq:g2}
}
\end{figure}

A detailed model of the pair production process must include a more
careful description of the collision geometry of colliding and expanding condensates
as well the effect of the condensates' mean field on the scattered atoms, something
which is neglected in the above discussion.
A rough estimate of the mean field effect is found by adding
the chemical potential to the kinetic energy of a scattered atom. 
This gives an additional velocity broadening of order $0.03\,v_{\rm rec}$, not
entirely negligible compared to the observed widths. 
Several workers are developing such models.
The correlation functions we observe lend themselves to an investigation of 
Cauchy-Schwartz inequalities \cite{walls:91}.
A cross correlation (back to back) greater than an autocorrelation (collinear) violates
a Cauchy-Schwartz inequality for classical fields. 
Sub-Poissonian number differences between opposite directions should also be present
 \cite{savage:06}.
A future publication will discuss these aspects of the experiment.

\begin{acknowledgments}
We acknowledge valuable discussions with K. Kheruntsyan, K. M\o{}lmer and M. Trippenbach. Our group is a member of the IFRAF institute, and is supported by the french ANR, the QUDEDIS program and the SCALA program of the European Union.
\end{acknowledgments}

%\bibliographystyle{prsty}

%\bibliography{apspairs}% Produces the bibliography via BibTeX.

\begin{thebibliography}{10}
\bibitem{scatlength:07}
We extrapolate from \cite{moal:06,leo:01} a value of $5.3$~nm for the scattering length between $m=0$ atoms.
\end{thebibliography}

\end{document}